\documentclass[twoside,twocolumn,9pt]{article}
\usepackage{extsizes}
\usepackage[super,sort&compress,comma]{natbib} 
\usepackage[version=3]{mhchem}
\usepackage[left=1.5cm, right=1.5cm, top=1.785cm, bottom=2.0cm]{geometry}
\usepackage{balance}
\usepackage{mathptmx}
\usepackage{sectsty}
\usepackage{graphicx} 
\usepackage{lastpage}
\usepackage[format=plain,justification=justified,singlelinecheck=false,font={stretch=1.125,small,sf},labelfont=bf,labelsep=space]{caption}
\usepackage{float}
\usepackage{fancyhdr}
\usepackage{fnpos}
\usepackage[english]{babel}
\addto{\captionsenglish}{%
  
}
\usepackage{array}
\usepackage{droidsans}
\usepackage{charter}
\usepackage[T1]{fontenc}
\usepackage[usenames,dvipsnames]{xcolor}
\usepackage{setspace}
\usepackage[compact]{titlesec}
\usepackage{hyperref}

\usepackage{epstopdf}

\usepackage{physics}

\definecolor{cream}{RGB}{222,217,201}
\newcommand{\red}[1]{#1}

\begin{document}

\pagestyle{fancy}
\thispagestyle{plain}
\fancypagestyle{plain}{
\renewcommand{\headrulewidth}{0pt}
}

\makeFNbottom
\makeatletter
\renewcommand\LARGE{\@setfontsize\LARGE{15pt}{17}}
\renewcommand\Large{\@setfontsize\Large{12pt}{14}}
\renewcommand\large{\@setfontsize\large{10pt}{12}}
\renewcommand\footnotesize{\@setfontsize\footnotesize{7pt}{10}}
\makeatother

\renewcommand{\thefootnote}{\fnsymbol{footnote}}
\renewcommand\footnoterule{\vspace*{1pt}%
\color{cream}\hrule width 3.5in height 0.4pt \color{black}\vspace*{5pt}} 
\setcounter{secnumdepth}{5}

\makeatletter 
\renewcommand\@biblabel[1]{#1}            
\renewcommand\@makefntext[1]%
{\noindent\makebox[0pt][r]{\@thefnmark\,}#1}
\makeatother 
\renewcommand{\figurename}{\small{Fig.}~}
\sectionfont{\sffamily\Large}
\subsectionfont{\normalsize}
\subsubsectionfont{\bf}
\setstretch{1.125} 
\setlength{\skip\footins}{0.8cm}
\setlength{\footnotesep}{0.25cm}
\setlength{\jot}{10pt}
\titlespacing*{\section}{0pt}{4pt}{4pt}
\titlespacing*{\subsection}{0pt}{15pt}{1pt}

\fancyfoot{}
\fancyfoot[LO,RE]{\vspace{-7.1pt}\includegraphics[height=9pt]{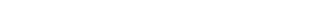}}
\fancyfoot[CO]{\vspace{-7.1pt}\hspace{13.2cm}\includegraphics{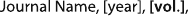}}
\fancyfoot[CE]{\vspace{-7.2pt}\hspace{-14.2cm}\includegraphics{head_foot/RF}}
\fancyfoot[RO]{\footnotesize{\sffamily{1--\pageref{LastPage} ~\textbar  \hspace{2pt}\thepage}}}
\fancyfoot[LE]{\footnotesize{\sffamily{\thepage~\textbar\hspace{3.45cm} 1--\pageref{LastPage}}}}
\fancyhead{}
\renewcommand{\headrulewidth}{0pt} 
\renewcommand{\footrulewidth}{0pt}
\setlength{\arrayrulewidth}{1pt}
\setlength{\columnsep}{6.5mm}
\setlength\bibsep{1pt}

\makeatletter 
\newlength{\figrulesep} 
\setlength{\figrulesep}{0.5\textfloatsep} 

\newcommand{\topfigrule}{\vspace*{-1pt}%
\noindent{\color{cream}\rule[-\figrulesep]{\columnwidth}{1.5pt}} }

\newcommand{\botfigrule}{\vspace*{-2pt}%
\noindent{\color{cream}\rule[\figrulesep]{\columnwidth}{1.5pt}} }

\newcommand{\dblfigrule}{\vspace*{-1pt}%
\noindent{\color{cream}\rule[-\figrulesep]{\textwidth}{1.5pt}} }

\makeatother

\twocolumn[
  \begin{@twocolumnfalse}
{\includegraphics[height=30pt]{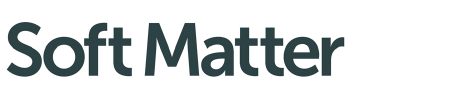}\hfill\raisebox{0pt}[0pt][0pt]{\includegraphics[height=55pt]{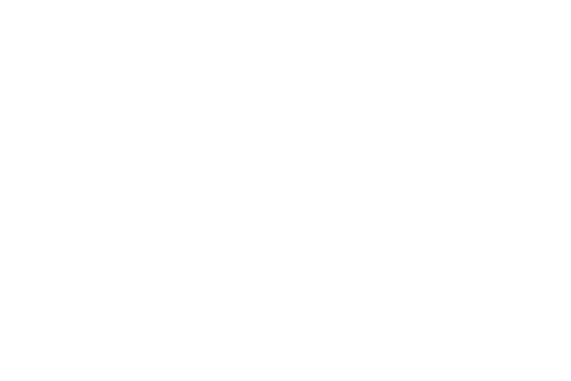}}\\[1ex]
\includegraphics[width=18.5cm]{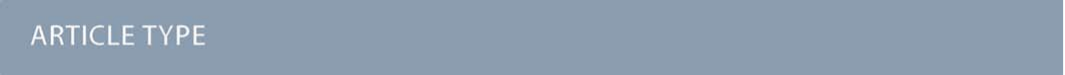}}\par
\vspace{1em}
\sffamily
\begin{tabular}{m{4.5cm} p{13.5cm} }

\includegraphics{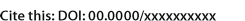} & \noindent\LARGE{\textbf{Splay and polar order in a system of hard \red{pear-like molecules: confrontation of Monte Carlo numerical simulations with density functional theory calculations}}} \\
\vspace{0.3cm} & \vspace{0.3cm} \\

 & \noindent\large{Piotr Kubala,$^{\ast}$\textit{$^{a}$} and Micha\l{} Cie\'sla\textit{$^{b}$}} \\

\includegraphics{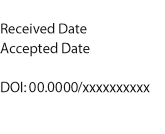} & \noindent\normalsize{Recent experimental discoveries of novel nematic types with polar order, including ferroelectric nematic and splay nematic have brought the resurgence of the interest in  polar and modulated phases. One of the most important factors that is widely believed to be crucial for the formation of the new phases is the \red{pear-like} shape of the mesogenic molecules. Such \red{molecules} were treated using second-virial density functional theory in [De Gregorio, P \textit{et al.}, \textit{Soft Matter}, 2016, \textbf{12(23)}, 5188-5198], where the authors showed that the $K_{11}$ splay elastic constant can become negative due to solely entropic reasons leading to long-range splay and polar correlations. To verify whether the predictions are correct, we performed Monte Carlo simulations of the same hard-core \red{molecules} used in the DFT study. As our results suggest, no \red{polar or modulated} liquid crystalline phases emerge; polar and splay correlations are at most short-range or completely absent. On the other hand, a polar ferroelectric \red{splay} crystal was observed. 
} \\

\end{tabular}

 \end{@twocolumnfalse} \vspace{0.6cm}
  ]

\renewcommand*\rmdefault{bch}\normalfont\upshape
\rmfamily
\section*{}
\vspace{-1cm}

\footnotetext{\textit{$^{a}$~Institute of Theoretical Physics, Jagiellonian University in Krak\'ow, \L{}ojasiewicza 11, 30-348 Krak\'ow, Poland. Email: piotr.kubala@doctoral.uj.edu.pl}}

\footnotetext{\textit{$^{b}$~Institute of Theoretical Physics, Jagiellonian University in Krak\'ow, \L{}ojasiewicza 11, 30-348 Krak\'ow, Poland. Email: michal.ciesla@uj.edu.pl}}



\section{Introduction}

\begin{figure*}[htb]
    \centering
    \includegraphics[width=0.95\linewidth]{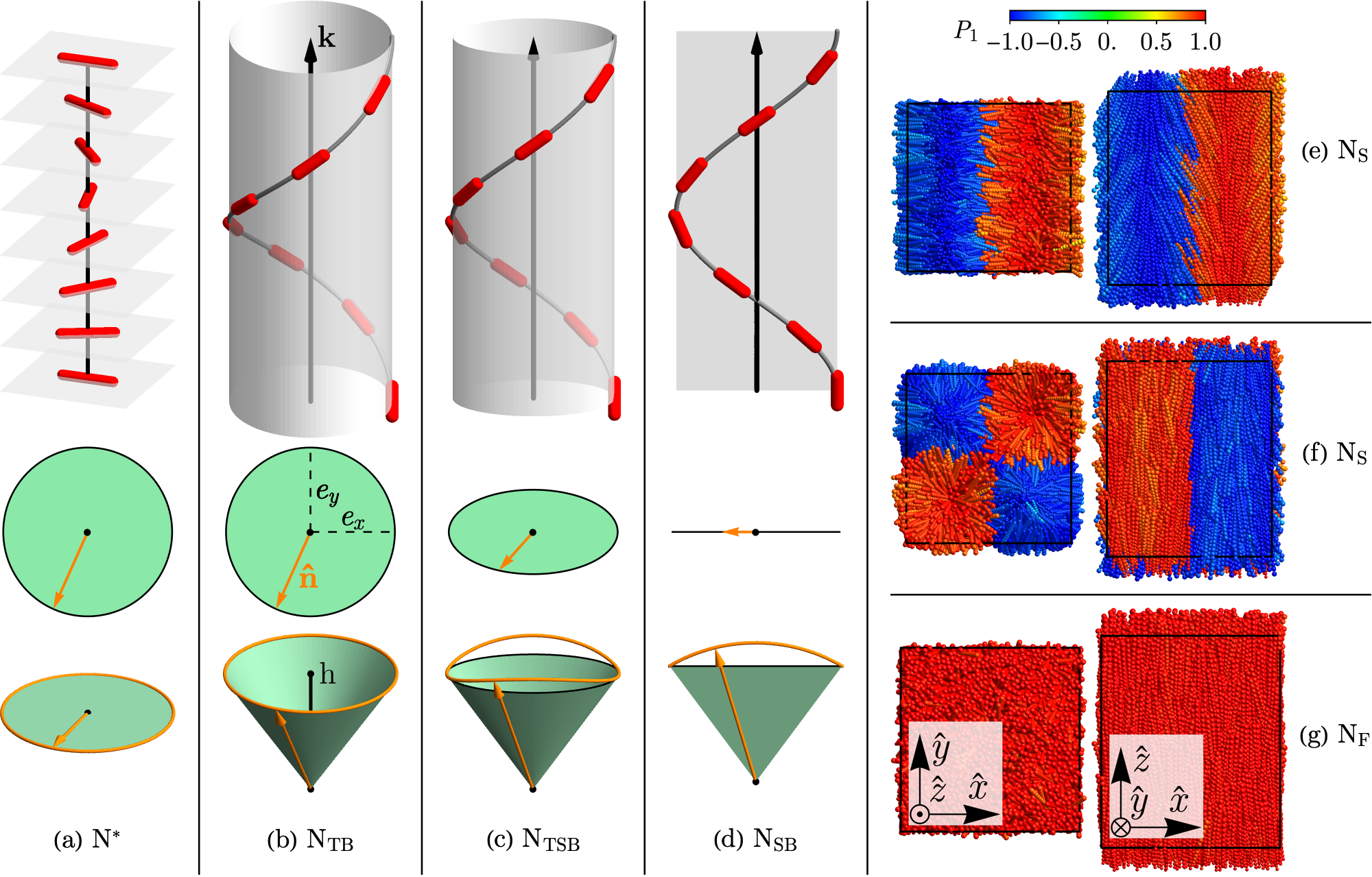}
    \caption{\red{Schematic representations of selected modulated phases: (a) cholesteric $\text{N}^\text{*}$, (b) twist-bend nematic $\text{N}_\text{TB}$, (c) twist-splay-bend nematic $\text{N}_\text{TSB}$, (d) splay-bend nematic $\text{N}_\text{SB}$, (e) single splay nematic $\text{N}_\text{S}$, (f) double splay nematic $\text{N}_\text{S}$, (g) ferroelectric nematic $\text{N}_\text{F}$. In phases from columns (a)-(d) the director $\vu{n}$ precesses around an asymmetric cone with axes $e_x$ and $e_y$ and height $h$ along the modulation wavevector $\vb{k} \parallel \vu{z}$: (a) $e_x = e_y$, $h = 0$, (b) $e_x = e_y$, $h \neq 0$, (c) $e_x \neq e_y$, $h \neq 0$, (d) $e_x \neq 0$, $e_y = 0$, $h \neq 0$. Within each column, the top graphics illustrates how the preferred direction of molecules changes along $\vb{k}$, the center one show the top view of the precession cone and the bottom one -- the perspective view of the cone. Panels (e), (f) show two hypothetical realization of splay nematic: (e) single splay with stripped pattern and (f) double splay with checkerboard pattern, while (g) is the ferroelectic nematic with spontaneous macroscopic polarization. Within each row, left panel is the top view, while the right panel is the side view. The color-coding is according to $P_1 = \vu{a} \cdot \vu{z}$, where $\vu{a}$ is the main molecular axis. Please note that all presented phases may theoretically occur in nematic and smectic variants depending on whether density modulation is present or not.}}
    \label{fig:fauna}
\end{figure*}

Since their discovery in 1888\cite{Reinitzer1888}, liquid crystals have become one of the most fruitful areas of study for various phases characterized by unexpected properties resulting from their internal structure based on the orientational order of the anisotropic molecules that built them \cite{Stephen1974, Chandrasekhar1992, deGennes1993}. The internal structure of liquid crystals leads to their anisotropic macroscopic properties, which are the basis for numerous applications, the most popular of which are liquid crystal displays (LCD-s) \cite{Uchida2022}. 

The most common and simplest liquid crystalline phases are nematics and smectics, where the director $\vu{n}$ -- the direction along which \red{the main axes of the molecules} tend to align -- is the same in the whole system. Apart from these structures, modulated phases, where the direction of the ordering changes, are of particular interest. Such phases are typically induced by the breaking of specific symmetry at the molecular level, and therefore, instead of parallel alignment of neighboring molecules, a slight tilt in their orientations is preferred; thus, the director is no longer spatially constant. In general, the deformations of a uniform director field are usually described in terms of the Oseen-Zocher-Frank free energy\cite{Oseen1933,Zocher1933,Frank1958}.
\begin{equation}
    \mathcal{F}_\text{OZF} = \frac{1}{2} K_{11} [\vu{n} (\div{\vu{n}})]^2 + \frac{1}{2} K_{22} [\vu{n} \vdot (\curl{\vu{n}})]^2 + \frac{1}{2} K_{33}[\vu{n} \cp (\curl{\vu{n}})]^2.
\end{equation}
The subsequent terms correspond to, respectively, splay, twist, and bend deformation modes. $K_{ii}$, called the elastic constants, determine the energetic cost of deviation from a uniform director field. In most scenarios, they are all positive with $K_{22} < K_{11} < K_{33}$\cite{Demus2008,Luckhurst2001}, thus any deformations are opposed by restoring torques. There are, however, experimental, theoretical, and numerical cases, where one of them is anomalously low, or even negative \cite{Meyer1976,Dozov2001,Memmer2002,Shamid2013,Borshch2013,Chen2013,Greco2015,Allesandro2017,Chiappini2021,Kubala2022}. In the latter case, the uniform nematic \red{or smectic} ceases to be the stable structure in favor of modulated phases \red{(see Fig.~\ref{fig:fauna})}. These include cholesterics \cite{Harris1999}, twist-bend nematic $\text{N}_\text{TB}$\cite{Borshch2013,Chen2013,Greco2014,Chiappini2021,Kubala2022}, splay nematic $\text{N}_\text{S}$\cite{Dhakal2010,Mertelj2018,Rosseto2020,Sebastian2020}, splay-bend nematic $\text{N}_\text{SB}$\cite{Archbold2015,Pajak2018,Chaturvedi2019,Fernandez2020} and smectic $\text{Sm}_\text{SB}$\cite{Chiappini2021,Kubala2022} as well as splay-twist-bend smectic $\text{Sm}_\text{STB}$\cite{Chiappini2021}. Of particular interest has recently been splay nematic $\text{N}_\text{S}$ phase, as it is closely related to the ferroelectric nematic $\text{N}_\text{F}$\cite{Mandle2017,Mertelj2018,Chen2020,Sebastian2022} with global polarization, which has been gaining a lot of traction due to its scientific and practical significance\cite{Chen2020}.

Among the most important factors responsible for the softening of the elastic constants is the shape effect. It was repeatedly proven using theoretical models\cite{Greco2015,Gregorio2016,Chiappini2021} as well as molecular dynamics (MD) and Monte Carlo (MC) simulations\cite{Memmer2002,Greco2015,Chiappini2021,Kubala2022}, that purely repulsive bent-core (banana-shaped) \red{molecules} spontaneously form $\text{N}_\text{TB}$ phase with broken mirror symmetry. One of these works is that of De Gregorio \emph{et al.}\cite{Gregorio2016}, which, by means of density functional theory (DFT) calculations, predicts the existence of the $\text{N}_\text{TB}$ phase for banana-shaped \red{molecules}, as well as the splay nematic $\text{N}_\text{S}$ phase for \red{pear-like molecules}. Moreover, the onset of spontaneous director field modulation was accompanied by the polar order; in the former case, the bend was coupled to the transversal polarization of the \red{molecule}, while in the latter case, it was coupled to the longitudinal polarization. To our knowledge, the $\text{N}_\text{S}$ phase was not observed in an MD or MC study with purely entropic interactions. 
\red{On the other hand, polar nematic and polar smectic were observed in MC simulations in the system built of molecules combining two rigidly connected centers interacting by Gay-Berne \cite{Gay1981} and Lennard-Jones potentials. Additionally, it appeared that the additional dipole-dipole interaction does not influence the existence of the polar phases \cite{Berardi2001, Barmes2003}. However, it is known that the Gay-Berne potential with a dipole at one end of the molecule leads to formation of the bilayer smectic phase \cite{Houssa2009}.}

\begin{figure}[htbp]
    \centering
    \includegraphics[width=0.8\linewidth]{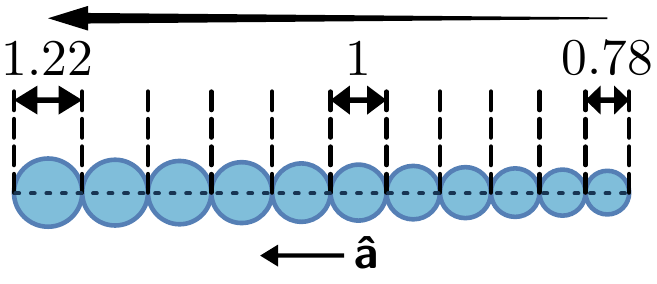}
    \caption{Illustration of a \red{pear-like} molecule used in the study. It consists of eleven co-linear tangent balls with diameters increasing linearly from 0.78 to 1.22. Long molecular axis $\vu{a}$ points from the smallest to the largest ball. The same model was used in Ref.~\cite{Gregorio2016}.}
    \label{fig:wedge}
\end{figure}

In this manuscript, we attempt to recreate the splay nematic or the splay smectic phase using MC simulations of the hard-core \red{pear-like} model (see Fig.~\ref{fig:wedge}). The \red{molecules} are built of eleven co-linear tangent hard spheres with diameters linearly increasing from 0.78 to 1.22. The same model was used in Ref.~\cite{Gregorio2016}, where the negative $K_{11}$ splay constant was reported. A similar \red{molecule}, with six beads instead of eleven, was the topic of our previous work\cite{Kubala2022phases}. As our results suggest, long-range polar and splay order appear only in the crystalline phase, whereas in the nematic and smectic phase, they are present at most locally.

The paper is structured as follows. In Sec.~\ref{sec:mc} we describe the numerical methods that we were using. Then, in Secs.~\ref{sec:order_params} and \ref{sec:corr} we describe the order parameters and \red{pair-}correlation functions. Sec.~\ref{sec:phases} walks through all phases, from the isotropic liquid to the solid state. Finally, Sec.~\ref{sec:polar_splay} discusses the presence and range of polar and splay correlations in the system, while Sec.~\ref{sec:conclusions} sums up the findings and outlines the potential further directions of studies. \red{Additionally, Appendix~\ref{sec:equilib} shows the equilibration curves of the packing fraction and selected order parameters.}

\section{Methods}

\subsection{Monte Carlo simulations} \label{sec:mc}

To establish the phase sequence for the \red{molecule} model under consideration, we performed NpT Monte Carlo simulations\cite{Wood1968,Allen2017,Allen2019} of $N = 4320$ \red{molecules} using our RAMPACK software\footnote{\url{https://github.com/PKua007/rampack}}. As the interactions are purely hard-core, the only independent thermodynamic parameter is the reduced pressure and temperature ratio $p^*/T^* = p V/k_B T$, where $V$ is the volume of a single \red{molecule} and $k_B$ is the Boltzmann constant. We investigated $p^*/T^*$ from the range $[1.0, 11.0]$ corresponding to packing fractions $\eta \in [0.15, 0.51]$, which covered the entire phase sequence, from the isotropic liquid to the solid state. \red{For each $p^*/T^*$ value, we performed a single simulation run consisting of the equilibration phase and the production phase.} Both phases lasted around $10^8$ Monte Carlo cycles \red{for most $p^*/T^*$ values, however the crystalline phase required extending the equilibration phase to $3 \cross 10^8$ cycles}. \red{In the production phase, instantaneous order parameters and values of \red{pair-}correlation functions were computed every $10^5$ cycles and averaged at the end}. Each cycle consisted of $N$ rototranslation moves, $N/10$ flip moves, and a single box move. In the rototranslation move, a molecule was chosen at random and its position and rotation were perturbed randomly. During the flip move, a random \red{molecule} was rotated 180 degrees so that the sense of the molecular axis $\vu{a}_i$ changed ($\vu{a}_i \rightarrow -\vu{a}_i$), while the direction remained unchanged. The flip move facilitated polar order relaxation, especially for high packing fractions. For both types of moves, perturbations were accepted only if no overlap was introduced. During the box move, box vectors $\vb{b}_1, \vb{b}_2, \vb{b}_3$ were randomly perturbed \red{and the positions of particles' centers were appropriately updated
}. Overlapping configurations were rejected immediately, while non-overlapping ones were accepted with a probability given by the Metropolis\red{-Wood} criterion \red{\cite{Wood1968jcp,Wood1968}}.
\begin{equation}
    P = \min\qty{1, \exp(N \log \frac{V_1}{V_0} - \frac{p \Delta V}{T})},
\end{equation}
where $V_0 = \abs{\vb{b}_1 \vdot (\vb{b}_2 \cp \vb{b}_3)}$ is the box volume before the move, $V_1$ -- after the move, and $\Delta V = V_1 - V_0$. For $p^*/T^* < 9.0$ ($\eta < 0.45$), the box was tetragonal and its dimensions were logarithmically scaled, with the $z$ axis perturbed independently of the $x$ and $y$ axes (making the scaling anisotropic). For $p^*/T^* \ge 9.0$ ($\eta \ge 0.45$), the box was triclinic and the box vectors were altered by adding small random vectors to them. Initial configurations for all liquid phases were slightly diluted \red{($\eta \approx 0.37$)} hexagonal honeycomb layers with random up-down orientations of \red{molecules}. For the crystalline phase at $p^*/T^* = 11$ ($\eta = 0.51$), the initial configuration was the final snapshot of the $p^*/T^* = 8$ ($\eta = 0.44$) smectic A liquid. In all the cases periodic boundary conditions were used.

\subsection{Order parameters} \label{sec:order_params}

To quantitatively characterize the observed phases, we ensemble averaged three order parameters: nematic order $\expval{P_2}$, smectic order $\expval{\tau}$ and hexatic bond order $\expval{\psi_6}$. The nematic order parameter $P_2$ is defined as\cite{vieillard1974, Eppenga1984}
\begin{equation} \label{eq:p2}
    P_2 = \frac{3}{2}\qty(\vu{a} \cdot \vu{n} - \frac{1}{3}),
\end{equation}
where $\vu{a}$ is a molecular axis and $\vu{n}$ is the director. The mean $P_2$ in a single snapshot can be conveniently computed using the $\vb{Q}$-tensor\cite{Eppenga1984}:
\begin{equation} \label{eq:Q}
    \vb{Q} = \frac{1}{N} \sum_{i=1}^{N} \frac{3}{2}\qty(\vu{a}_i \otimes \vu{a}_i - \frac{1}{3} \vb{I}),
\end{equation}
where the summation goes over all \red{molecules} in the system\red{, and $\vb{I}$ denotes unity matrix}. In this formulation, $P_2$ is the eigenvalue of $\vb{Q}$ with the highest magnitude, and $\vu{n}$ is the corresponding eigenvector. Then $\expval{P_2}$ is calculated by averaging it over uncorrelated system snapshots.

The smectic order parameter can be defined as\cite{deGennes1993}
\begin{equation}
    \expval{\tau} = \expval{\frac{1}{N}\abs{\sum_{i=1}^{N}\exp(\imath \vb{k} \vdot \vb{x}_i)}},
\end{equation}
where $\vb{k}$ is the smectic wavevector, $\vb{x}_i$ is the position of the $i$-th \red{molecule}, $\expval{\dots}$ is ensemble averaging, and $\imath$ is imaginary unit. In a finite system, $\vb{k}$ must be compatible with periodic boundary conditions. In general, one can use the following formula to enumerate all possibilities:
\begin{equation}
    \vb{k} = h \vb{g}_1 + k \vb{g}_2 + l \vb{g}_3.
\end{equation}
Here, $\vb{g}_i$ are reciprocal box vectors\cite{Kittel2018} and $h, k, l$ are integers (the Miller indices). In this study, all initial states had four layers stacked along the $z$-axis, therefore, we assumed $hkl = 004$.

To quantify local hexatic order, one can use the hexatic bond order parameter $\expval{\psi_6}$\cite{Nelson2012}. It is essentially a two-dimensional parameter; thus, its computation has to be restricted to a single plane in a three-dimensional system. A natural choice is to compute it for each layer separately and average it over all layers. For a single layer $l$ it is defined as
\begin{equation}
    \psi_6^l = \frac{1}{N_l} \sum_{i=1}^{N_l}\frac{1}{6} \abs{\sum_{j \in \text{6NN}(i)} \exp(6 \imath \theta_{ij})},
\end{equation}
where $N_l$ is the number of \red{molecules} in the $l$-th layer, $\text{6NN}(i)$ is a list of the six nearest neighbors of the $i$-th \red{molecule} and $\theta_{ij}$ is an angle between the projection of vector joining $i$-th and $j$-th \red{molecule} onto the layer and an arbitrary direction within the layer. Then, $\psi_6^l$ is averaged over all four layers and uncorrelated system snapshots
\begin{equation}
    \expval{\psi_6} = \expval{\frac{1}{4} \sum_{l=1}^{4} \psi_6^l}.
\end{equation}
\red{The maximal value $\expval{\psi_6} = 1$ is achieved for a perfect hexagonal arrangement, while for a set of random points $\expval{\psi_6} \approx 0.37$. As the computation of $\expval{\psi_6}$ requires well-defined layers, it will be quantified only for the density-modulated (smectic and crystalline) phases.}

\red{To measure local polarization one can define the following quantity
\begin{equation}
    \expval{P_{1}^{(r)}} = \expval{\frac{1}{N}\sum_{i=1}^{N} \frac{1}{r+1} \abs{ \sum_{j \in r\text{NN},i} \vu{n}_i \vdot \vu{a}_j}}.
\end{equation}
Here, ($r\text{NN},i$) is a set containing particle $i$ and its $r$ nearest neighbors, $\vu{n}_i$ is local director computed for this set using the $\vb{Q}$ tensor \eqref{eq:Q}. Locality can be controlled by the rank $r$ -- we selected $r = 9$. $\expval{P_{1}^{(9)}}$ reaches 1 when molecules' polarizations are identical. Due to taking the modulus, the value of $\expval{P_{1}^{(9)}}$ for random polarizations is non-zero and depends on $r$; for $r = 9$ it is $\expval{P_{1}^{(9)}} \approx 0.17$. For our convenience, in next parts of the manuscript we will write $\expval{P_{1}}$ without the rank.

}

\subsection{Correlations} \label{sec:corr}

As postulated in Ref.~\cite{Gregorio2016}, splay deformation mode is coupled to a longitudinal polarization of the molecule. Thus, we should be looking for a long-range order in the polarization field and splay correlations. To probe \red{polarization correlations}, we used the transversal $S_\perp^{110}(r_\perp)$ \red{pair-}correlation function\cite{Stone1978}. It is defined layer-wise, similar to $\psi_6$:
\begin{equation} \label{eq:s110}
    S_\perp^{110}(r_\perp) = \expval{\frac{1}{4} \sum_{l=1}^{4} \expval{\vu{a}_{i_l} \cdot \vu{a}_{j_l}}_{i_l j_l}},
\end{equation}
where $l$ is the layer number, $\expval{\dots}_{i_l j_l}$ denotes averaging over all pairs $(i_l, j_l)$ of \red{molecules within the $l$-th layer}, whose transversal distance $r_\perp$ lies in the range $[r_\perp - \dd{r}, r_\perp + \dd{r}]$ and $2\dd{r}$ is the numerical bin size. \red{The transversal distance $r_\perp$ is calculated along the layers -- more precisely, it is the length of the projection of the vector $\vb{x}_{ij} = \vb{x}_j - \vb{x}_i$ joining \red{molecules} $i$ and $j$ onto the nearest layer. In the nematic phase with no density modulation, $r_\perp$ could be computed by projecting all \red{molecules} onto a single plane. However, in order to facilitate capturing local correlations, we divide nematic snapshots arbitrarily into four layers, whose width is similar to the \red{molecule}'s length and use the same formula \eqref{eq:s110} as for modulated phases.}

Quantifying splay deformation is significantly more difficult. The reason is that, contrary to all the observables introduced earlier, the splay term is a derivative of the director field itself. Thus, to compute it numerically, a sufficiently smooth vector field estimation is needed. This requires ensemble averaging of the field prior to computation. If a long-range order is not present, instantaneous short-range correlations evolve with time, which, with the help of translational Goldstone mode, average out to a uniform, nematic-like director field. In order to probe local splay correlations, we propose a different scheme. The director field with only the splay deformation and a single singularity in the origin is
\begin{equation} \label{eq:n_splay}
    \vu{n}(\vb{R}) = \frac{\vb{R}}{\norm{\vb{R}}},
\end{equation}
which can be described as a hedgehog-like structure. \red{Here, $\vb{R}$ is the vector joining hedgehog singularity with a given point.} If we choose a point \red{$\vu{n} = \vu{n}(\vb{R})$} and move in a transverse direction by \red{$\vb{R}_\perp$, $\vb{R}_\perp \perp \vb{R}$}, to \red{$\vu{n}' = \vu{n}(\vb{R} + \vb{R}_\perp)$}, the angle between $\vu{n}$ and $\vu{n}'$ should be approximately a linear function of \red{$\norm{\vb{R}_\perp}$}, as long as \red{$\norm{\vb{R}_\perp} \ll \norm{\vb{R}}$}. Based on this \red{and assuming that $\vb{R}_\perp$ is almost parallel to the layers}, we define $P(\theta|r_\perp)$ as a conditional probability distribution of finding two \red{molecules} $i$ and $j$ with a transversal distance $r_\perp$, whose molecular axes form angle $\theta = \cos^{-1}\abs{\vu{a}_i \cdot \vu{a}_j}$ (which respects $\vu{n} \leftrightarrow -\vu{n}$ director symmetry), normalized as
\begin{equation}
    \int_{0}^{\pi} P(\theta|r_\perp) \dd{\theta} = 1 \qquad \forall r_\perp.
\end{equation}
Then, local splay correlation should manifest itself as a set of maxima of $P(\theta|r_\perp)$ as a function of both $\theta$ and $r_\perp$, for which $\theta$ and $r_\perp$ are linearly dependent.

\section{Results}

\subsection{Phase sequence} \label{sec:phases}

\begin{figure}[htbp]
    \centering
    \includegraphics[width=0.97\linewidth]{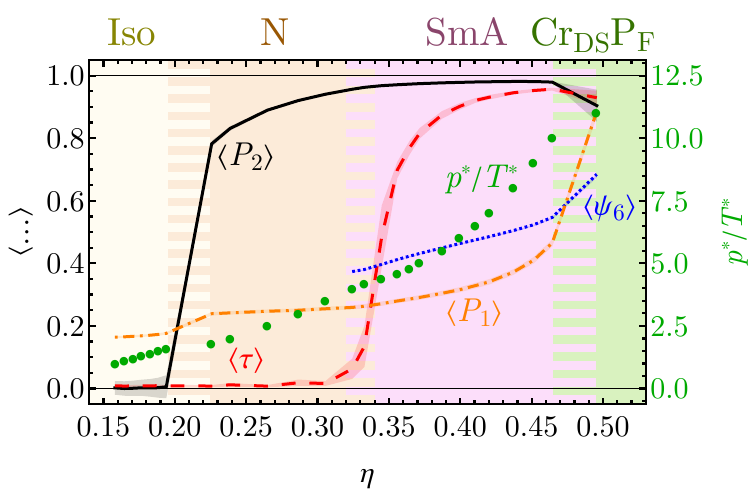}
    \caption{Phase sequence as a function of \red{the} packing fraction $\eta$. Colored areas represent the ranges of subsequent phases, as labeled above the diagram: Iso (isotropic), N (nematic), SmA (smectic A), $\text{Cr}_\text{DS}\text{P}_\text{F}$ (ferroelectric double-splay crystal). Lines are the order parameters' dependence on $\eta$: nematic order $\expval{P_2}$ (black solid), smectic order $\expval{\tau}$ (red dashed), hexatic bond order $\expval{\psi_6}$ (blue dotted) \red{and local polarization $\expval{P_1}$ (orange dot-dashed)}. \red{Green circles are the equation of state $p^*/T^*(\eta)$ -- each point corresponds to one simulated system. Left vertical axis labeled ``$\expval{\dots}$'' (black) is for order parameters, while the right one labeled ``$p^*/T^*$'' (green) -- for the equation of state.} Hatched regions indicate the vicinity of phase transitions. \red{Shaded areas around the lines denote sample standard deviations of the observables; for the majority of points, the errors are comparable with the width of the curves and thus barely visible.}}
    \label{fig:pd}
\end{figure}

\begin{figure*}[htbp]
    \centering
    \includegraphics[width=0.75\linewidth]{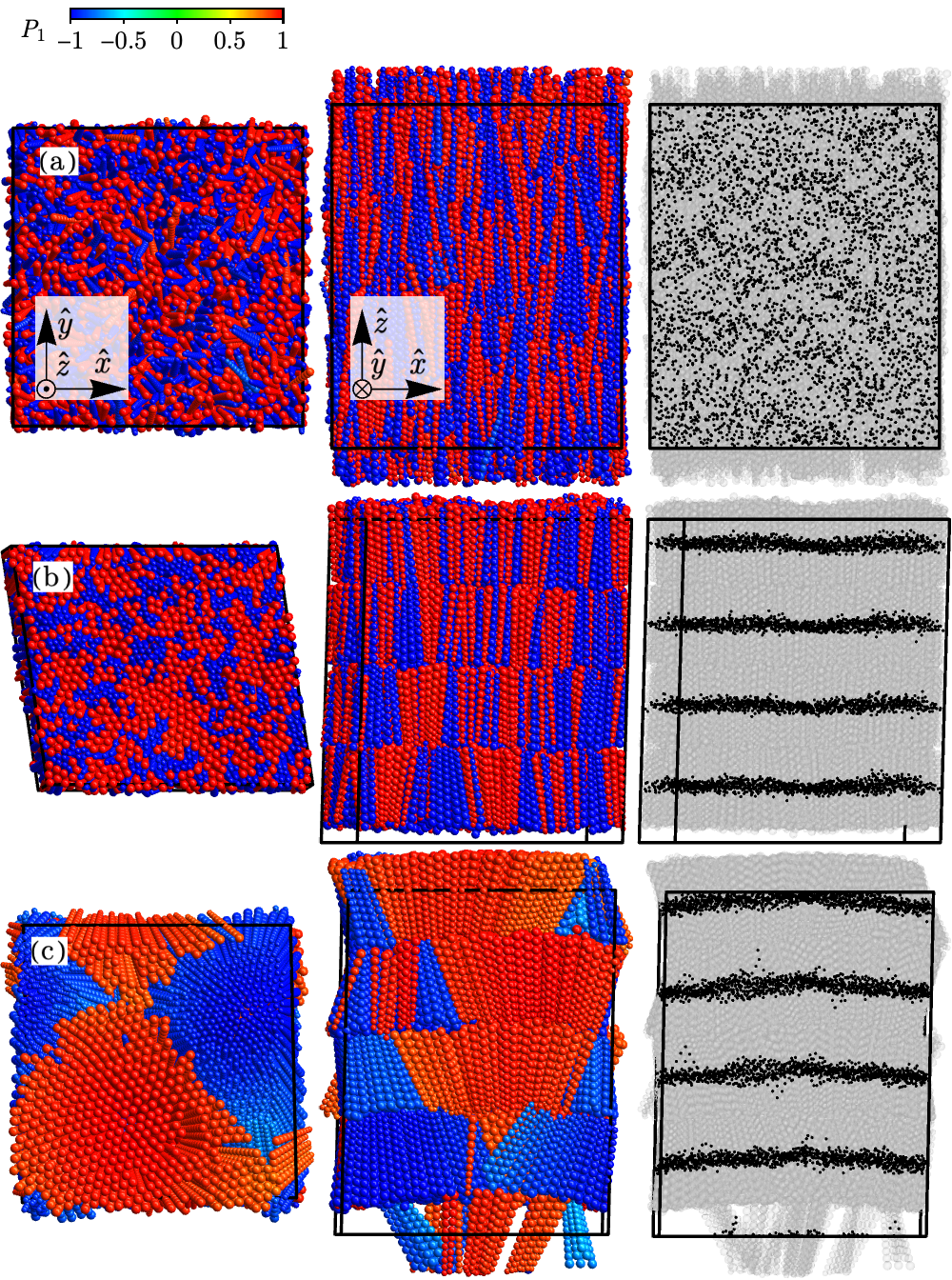}
    \caption{System snapshots for all observed phases apart from \red{the} isotropic liquid: (a) N ($\eta = 0.32$), (b) SmA ($\eta = 0.45$), (c) $\text{Cr}_\text{DS}\text{P}_\text{F}$ ($\eta = 0.51$). Each row depicts different aspects of a single snapshot. The first column is the top view of the simulation box, the second one -- the front view, and the last one shows the geometric centers (black dots) of \red{molecules} overlain on top of the side view (the second column). \red{Molecules} in the first two columns are color coded according to $P_1 = \vu{a}_i \vdot \vu{z}$.}
    \label{fig:pack}
\end{figure*}

Fig.~\ref{fig:pd} shows the phase sequence as well as the order parameters \red{and $p^*/T^*$ ratio} as a function of the packing fraction $\eta$ obtained from the Monte Carlo simulations. A rigorous determination of the orders of phase transitions is beyond the scope of this work, so the regions in the vicinity of transition points are shown with a hatch filling. The following phases were observed: isotropic (Iso), nematic (N), smectic A (SmA), and ferroelectric double-splay crystal ($\text{Cr}_\text{DS}\text{P}_\text{F}$). System snapshots for all phases, except Iso, are shown in Fig.~\ref{fig:pack}.

At the lowest densities, below $\eta = 0.2$, a disordered isotropic liquid is formed. It then undergoes a phase transition to the nematic phase [see Fig.~\ref{fig:pack}(a)], as indicated by the sharp jump in both the $\eta$ and the nematic order $\expval{P_2}$ parameter (black solid line in Fig.~\ref{fig:pd}). It may suggest that the transition is first order. No smectic order is observed, as indicated by the smectic $\expval{\tau}$ order parameter (red dashed line) -- it is close to zero in the entire range of the nematic phase. The nematic order increases monotonically from $\expval{P_2} = 0.8$ for $\eta = 0.23$ to $\expval{P_2} = 0.95$ for $\eta = 0.32$. Such high $\expval{P_2}$ values are above the typical range $\expval{P_2} \in [0.3, 0.7]$ seen in experiments\cite{Chandrasekhar1980}, but are not uncommon for hard-core systems treated numerically\cite{Vega2001,Lansac2003,Kubala2022,Kubala2022phases}. In particular, as seen in Fig.~\ref{fig:pack}(a) and discussed in detail in the next section, no long-range splay or polar order is present in the system. \red{Local polarization $\expval{P_1}$ is larger than in the Iso phase ($\expval{P_1} \approx 0.25$ compared to $\expval{P_1} \approx 0.17$) and rises very slowly with $\eta$. It, however, remains low in the whole range of the N phase.}

Around $\eta = 0.33$, an N-SmA phase transition occurs. The snapshot of the SmA phase can be seen in Fig.~\ref{fig:pack}(b). Nematic order $\expval{P_2}$ remains high and a rather quick ascent in $\expval{\tau}$ value to $\expval{\tau} = 0.8$ can be clearly seen. In experimental setups, the N-SmA phase transition can be both of the first and the second order \cite{Singh2000}, however, for idealized hard-core interactions the former one is usually observed\cite{McGrother1996,Polson1997,Lansac2003}. In our simulation data, $\eta$ did not experience a sudden jump and $\expval{\tau}$ rose fast, however smoothly, which would suggest a second-order phase transition. The smectic order $\expval{\tau}$ increases up to $\expval{\tau} = 0.94$ near the SmA-$\text{Cr}_\text{DS}\text{P}_\text{F}$ phase transition. At the same time, the hexatic bond order $\expval{\psi_6}$ rises almost linearly from $\expval{\psi_6} = 0.4$ at the lower phase boundary to $\expval{\psi_6} = 0.55$ at the top. A higher local hexatic order facilitates a more optimal packing, allowing the system to achieve a higher packing fraction $\eta$. On the other hand, $\expval{\psi_6} = 0.55$ is still lower than in systems with long-range hexatic order, where the values $\expval{\psi_6} > 0.7$ are observed\cite{Kubala2022phases}. Contrary to the nematic phase, the SmA system snapshot Fig.~\ref{fig:pack}(b) reveals some amounts of splay and polar order; however, it appears to be rather short-ranged. \red{Local polarization order is confirmed by $\expval{P_1}$ value which rises with the packing fraction $\eta$ and reaches $\expval{P_1} \approx 0.45$ near the SmA-$\text{Cr}_\text{DS}\text{P}_\text{F}$ phase boundary.}

Above $\eta = 0.46$, the system freezes and the ferroelectric double-splay crystal ($\text{Cr}_\text{DS}\text{P}_\text{F}$) phase emerges [cf. Fig.~\ref{fig:pack}(c)] with a sharp jump in $\eta$, $\expval{\psi_6}$ \red{and $\expval{P_1}$}, and, simultaneously, a slight decrease of $\expval{P_2}$ and $\expval{\tau}$. The phase was first observed and discussed in detail in Ref.~\cite{Kubala2022phases}, where a similar system of \red{pear-like molecules} was considered; however, the \red{molecules} were built of six instead of eleven beads. Here, we only give a brief summary. The phase is built up of layers. Within each layer, we observe square clusters of \red{molecules} with the same polarization, either ``up'' [$P_1 > 0$, colored red in Fig.~\ref{fig:pack}(c)] or ``down'' ($P_1 < 0$, colored blue). These clusters are arranged in a chessboard-like long-range pattern. \red{Local polarization in clusters is high with $\expval{P_1}$ reaching 0.89.} The \red{molecules} within each cluster form an ordered hexatic crystal with a high $\expval{\psi_6} = 0.7$ value. Long-ranged splay deformation is also clearly visible -- the \red{molecules} orient in a hedgehog-like pattern discussed in Sec.~\ref{sec:corr}. The presence of splay deformation weakens the global nematic and smectic order, as confirmed by a lowered $\expval{P_2} = 0.9$ and $\expval{\tau} = 0.92$ values, as compared to $\expval{P_2} = 0.97$ and $\expval{\tau} = 0.95$ in the dense SmA phase. Between the clusters, there are sharp domain walls, where the signs of both polarization and splay vector $\vu{n} (\div{\vu{n}})$ change abruptly. The ``chessboard'' layers are stacked on each other, matching the signs of polarization and splay deformation between them \red{-- as a consequence, polarization domains form long columns orthogonal to layers}\footnote{In contrast to antiferroelectric splay ($\text{Cr}_\text{S}\text{P}_\text{A}$) and double splay ($\text{Cr}_\text{DS}\text{P}_\text{A}$) crystal phases, \red{where the polarizations are alternating between the layers}, see Ref.~\cite{Kubala2022phases}}. However, it is important to note that more defects were observed \red{in} this system compared to \red{pear-like molecules} built of six balls in Ref.~\cite{Kubala2022phases}.

\subsection{Polar order and splay deformation} \label{sec:polar_splay}

\begin{figure}[htbp]
    \centering
    \includegraphics[width=0.9\linewidth]{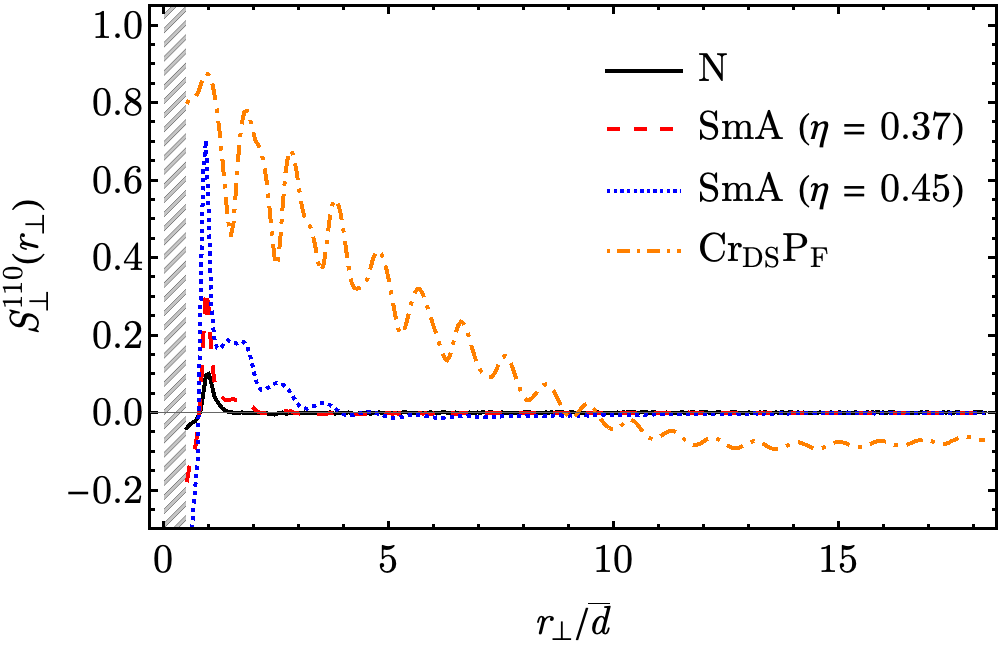}
    \caption{Layer-wise transversal \red{pair-}correlation $S_\perp^{110}(r_\perp)$ function for nematic (black solid line), low-density smectic A (red dashed line), high-density smectic A (blue dotted line) and ferroelectric double-splay crystal (orange dot-dashed line) as a function of transversal distance $r_\perp$. $\bar{d} = 1$ is the average diameter of balls building the \red{pear-like molecule}.}
    \label{fig:S110}
\end{figure}

\begin{figure}[htbp]
    \centering
    \includegraphics[width=0.65\linewidth]{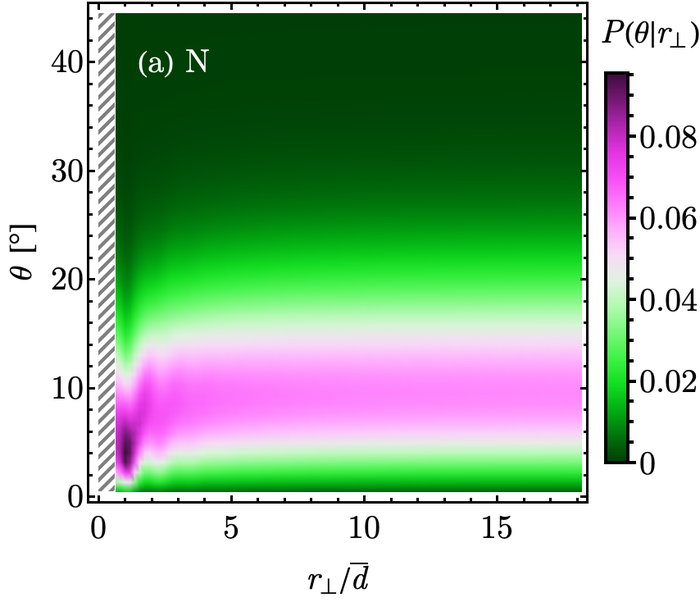}
    \includegraphics[width=0.65\linewidth]{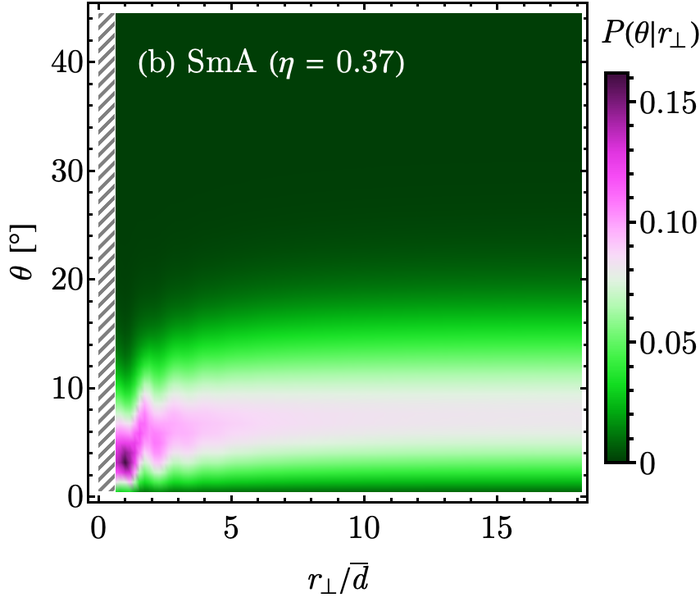}
    \includegraphics[width=0.65\linewidth]{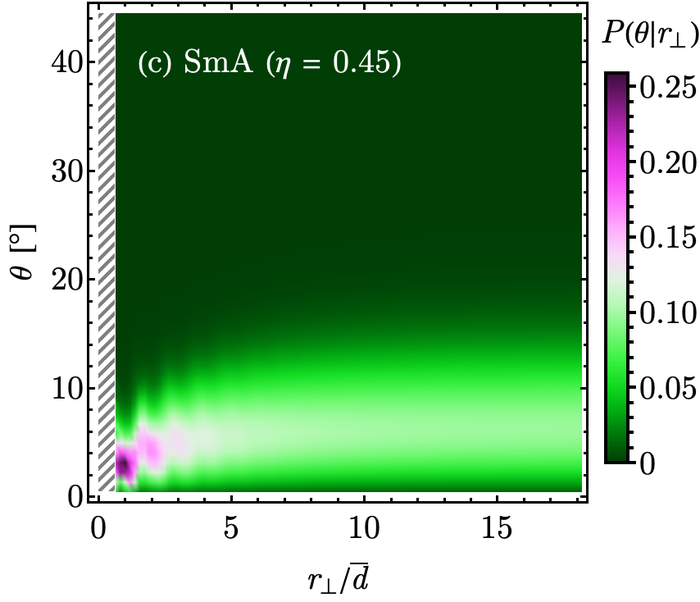}
    \includegraphics[width=0.65\linewidth]{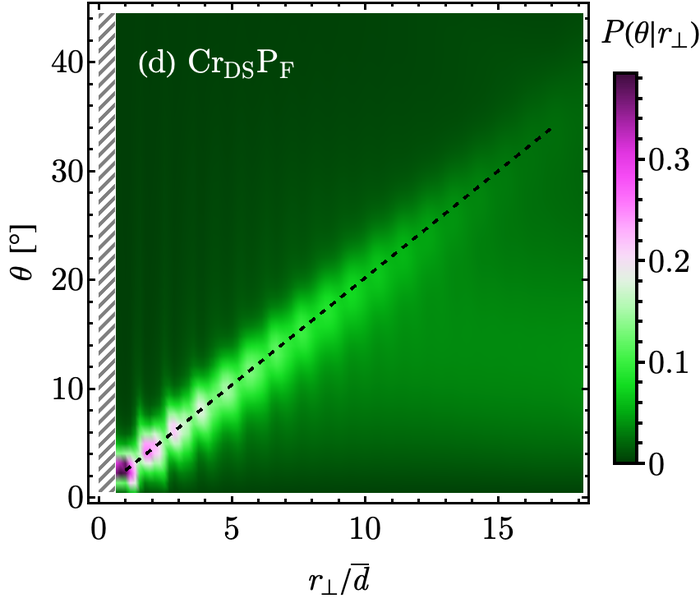}
    \caption{The evolution of the distribution of angles between the molecules with the transversal distance $r_\perp$ between them. For each section with a fixed value of $r_\perp$, the maps show a conditional probability density $P(\theta|r_\perp)$ of finding the two \red{molecules} with the angle given by the $y$-axis value. Subsequent panels correspond to different phases and packing fractions: (a) N ($\eta = 0.32$), (b) SmA ($\eta = 0.37$), (c) SmA ($\eta = 0.45$) and (d) $\text{Cr}_\text{DS}\text{P}_\text{F}$ ($\eta = 0.51$).}
    \label{fig:angle_hist}
\end{figure}

As noted in the Introduction, the authors in Ref.~\cite{Gregorio2016} predicted the softening of $K_{11}$ splay constant, when coupling with longitudinal polarization was taken into account. Thus, according to second-order DFT calculations, long-range splay and polar order should emerge. As there does not exist a non-singular director field with only splay deformation [see Eq.~\eqref{eq:n_splay} and Ref.~\cite{Mertelj2018}], it has to be accompanied by other deformations (twist, bend), or the singularities must be allowed. Rosseto \emph{et al.}\cite{Rosseto2020} discussed two types of splay nematic $\text{N}_\text{S}$\footnote{This structure has actually both splay and bend modes, however the name \emph{splay-bend} nematic $\text{N}_\text{SB}$ is widely used to describe a different director profile, \red{see Fig.~\ref{fig:fauna}(d) and Ref.\cite{Pajak2018}}.}, which can be observed. These structures are presented in Fig.~\ref{fig:fauna}(e,f). The first one, already known in the literature\cite{Chaturvedi2019}, was called \emph{single splay nematic} [Fig.~\ref{fig:fauna}(e)], where stripes with opposite splay and polarization are stacked in an alternating manner. In the lowest order, polarization and director fields are assumed to be
\begin{align}
    \vu{n}(x) &= (\sin[\Theta(x)], 0, \cos[\Theta(x)]), \qquad \Theta(x) = \Theta_0 \sin(\kappa x), \\
    \vb{P}(x) &= P_0 \cos(\kappa x) \vu{n}(x),
\end{align}
where $\Theta_0$ determines the strength of the deformation, $\kappa$ is the structure wavevector, and $P_0$ is the maximal local polarization. These profiles have been proven to work well near the critical point; however, for lower temperatures, sharper changes are observed\cite{Rosseto2020}. The second one was named \emph{double splay nematic} [Fig.~\ref{fig:fauna}(f)]. There, columns with opposite splay and polarization form a checkerboard pattern with the fields in lowest order given by
\begin{align}
    \vu{n}(x, y) &= \frac{(\Theta_0\sin(\kappa x)\cos(\kappa y), \Theta_0\cos(\kappa x)\sin(\kappa y), 1)}{\sqrt{\Theta_0^2\sin(\kappa x)^2\cos(\kappa y)^2 + \Theta_0^2\cos(\kappa x)^2\sin(\kappa y)^2 + 1}}, \\
    \vb{P}(x, y) &= 2P_0 \cos(\kappa x) \cos(\kappa y) \vu{n}(x).
\end{align}
\red{It is important to note that realization of such structures in a physical system would imply a severely hindered diffusion of molecules diffusion between ``up'' and ``down'' polarization domains as flipping molecule's polarization at a domain boundary requires passing a high energy barrier, which is a rare event.} As DFT calculations from Ref.~\cite{Gregorio2016} suggest, $K_{11}$ becomes negative for the number density $\rho \approx 0.050$ (see Fig.~3 therein), which corresponds to $\eta_S \approx 0.30$. In our simulations, it is the nematic phase near the N-SmA phase boundary. Thus, long-range polar and splay correlations should be observed for a high-density nematic and in the whole range of the smectic phase. As stated in Sec.~\ref{sec:phases} and visible in system snapshots, this is not the case. However, local correlations, especially in the smectic phase, seem to be present.

Polar order correlations can be quantified using $S_\perp^{110}(r_\perp)$. The dependence for selected packing fractions is shown in Fig.~\ref{fig:S110}. For a high-density nematic just below N-SmA transition point ($\eta = 0.32$, black solid line), lying above $\eta_S$, polar correlations are marginal -- they are non-zero only for nearest neighbors ($r_\perp \approx \bar{d}$, where $\bar{d} = 1$ is mean ball diameter) with $S_\perp^{110} \approx 0.15$. Slightly higher correlations are observed for a low-density smectic A ($\eta = 0.37$, red dashed line). Here, the maximum is $S_\perp^{110}(\bar{d}) \approx 0.3$ and the correlations have a slightly longer range, reaching $r_\perp \approx 2\bar{d}$. For high-density smectic A, the correlations are more prominent, with maximum $S_\perp^{110}(\bar{d}) \approx 0.7$, but still short-ranged, reaching $r_\perp \approx 4\bar{d}$. Local maxima lie near integer multiples of $\bar{d}$, which suggests that they correspond to nearest, next-nearest, next-next-nearest, etc. neighbors. Long-range correlations appear only after crystallization. For the $\text{Cr}_\text{DS}\text{P}_\text{A}$ phase at $\eta = 0.51$, positive correlations reach $S_\perp^{110} \approx 0.87$ and extend as far as $r_\perp \approx 9.5\bar{d}$. After that point, the polarizations are anticorrelated, which is in line with the checkerboard pattern visible in Fig.~\ref{fig:pack}(c). Interestingly, the phase realizes the double splay nematic structure proposed in Ref.~\cite{Rosseto2020}, but as a crystalline phase.

An insight into the range of splay order can be given by the relative probability density $P(\theta|r_\perp)$ of the angles $\theta$ between molecules at a given distance between them $r_\perp$. Histograms are shown in Fig.~\ref{fig:angle_hist}. In the nematic phase [Fig.~\ref{fig:angle_hist}(a)], $\theta \approx 10^\circ$ is the most probable angle for most $r_\perp$ with high spread. The highest maximum is for nearest neighbors ($r_\perp \approx \bar{d}$). \red{Additional local maxima are visible up to $r_\perp \approx 4\bar{d}$, which gives us the estimated range of angle correlations.}
It should be noted that \red{this} range is \red{slightly} higher than for the $S_\perp^{110}$ \red{pair-}correlation function. For a low-density smectic A [Fig.~\ref{fig:angle_hist}(b)], the histogram is qualitatively similar, with only a quantitative difference. The preferred angle is lower -- $\theta \approx 6^\circ$ with a narrower spread. Local maxima oscillations are also present with a similar range of $r_\perp \approx 4\bar{d}$. The situation changes slightly for a high-density smectic A [Fig.~\ref{fig:angle_hist}(c)]. Apart from a smaller preferred angle $\theta \approx 5^\circ$ with even less spread than for a low-density smectic, local maxima are \red{narrower} and their positions $(r_\perp, \theta)$ are close to a linear relation -- the signature we would expect from splay clusters. Unfortunately, the correlations are short-ranged, spanning up to $r_\perp \approx 5\bar{d}$. It changes drastically for the crystalline phase [Fig.~\ref{fig:angle_hist}(d)]. Maxima are clearly visible as far as $r_\perp \approx 15\bar{d}$ and they form an almost perfect linear relation given by $\theta = 0.53^\circ + 1.97^\circ r_\perp/\bar{d}$. The maxima become weaker with increasing $r_\perp$, which is expected since correlations between clusters with opposite polarization are also taken into account. These correlations are also the source of the wide local maximum around $\theta \approx 10^\circ$ for $r_\perp > 10\bar{d}$. Between the main maxima, one can observe long vertical lines. They likely also originate from off-lattice correlations between adjacent clusters.

As both the visual inspection of system snapshots and the quantitative \red{pair-}correlation functions clearly show that both long-range polar order and splay deformation are missing, while at the same time, the DFT \red{calculations predict} the N-$\text{N}_\text{S}$ transition, the question arises of why there is a mismatch between DFT \red{calculations} and MC simulations. The first possible reason, which also applies to most theoretical frameworks, is that DFT computations use a second-virial expansion with a correcting Parson-Lee factor\cite{Parsons1979,Lee1987}, while MC simulations do not restrict interactions to the two-particle term. Another reason, arguably more probable, is that the authors did not include density modulation in their calculations, while the alleged N-$\text{N}_\text{S}$ transition point is $\eta \approx 0.30$, which is close to the N-SmA transition point $\eta \approx 0.33$ in our simulations. As second-virial theories often give qualitatively correct results, however, with slightly wrong quantitative predictions\footnote{See for example Ref.~\cite{Greco2015} of the same research group, where theoretical and numerical N-$\text{N}_\text{TB}$ transition points are for number densities, respectively, $\rho \approx 0.056$ and $\rho \approx 0.050$.}, $\text{N}_\text{S}$ may have lower free energy than $N$ only over the N-SmA point, where non-modulated smectic A may be favorable over both modulated nematic and smectic phases. 

\section{Conclusions} \label{sec:conclusions}

We have performed Monte Carlo simulations of hard \red{pear-like molecules} built of eleven tangent balls. We scanned a wide range of packing fractions covering the whole phase sequence: isotropic (Iso), nematic (N), smectic A (SmA), and ferroelectric double-splay crystal ($\text{Cr}_\text{DS}\text{P}_\text{F}$). The phases were classified using visual inspection of simulation snapshots, order parameters, and \red{pair-}correlation functions. For the \red{molecule} under consideration, second-order DFT calculations~\cite{Gregorio2016} suggested the existence of a long-range polar and splay order for packing fractions $\eta > 0.30$, which cover high-density nematics and the entire smectic range. Despite the prediction of DFT \red{calculations}, no long-range polar or splay order was observed until the system crystallized. Only short-ranged correlations were present in the smectic phase, and in the nematic phase they were even less prominent. The possible reason why the theoretical predictions were not met may be that the theoretical N-$\text{N}_\text{S}$ transition point is close to the N-SmA point in our study, while at the same time, the smectic order was not included in the original manuscript. Another possibility is that the theory is only second-virial. However, since a negative value of the $K_{11}$ splay constant was proved to be theoretically possible, it is worth further exploring other variants of the \red{pear-like molecule} model, for example with a different length or a smooth, convex surface.

\section*{Data availability}
The datasets generated during and/or analyzed during the current study are available from P.K. upon reasonable request.

\section*{Code availability}

The source code of an original simulation package used to perform Monte Carlo sampling is available at \url{https://github.com/PKua007/rampack}.

\section*{Author Contributions}

P.K.: conceptualization, data curation, formal analysis, funding acquisition, investigation, software, visualization, writing. M.C.: conceptualization, writing.

\section*{Conflicts of interest}

There are no conflicts to declare.

\section*{Acknowledgements}

P.K. acknowledges the support of Ministry of Science and Higher Education (Poland) grant no. 0108/DIA/2020/49. M.C. acknowledges the support of National Science Center in Poland grant no. 2021/43/B/ST3/03135. The authors are grateful to Prof. Lech Longa for inspiring discussions. \red{A part of the numerical simulations was carried out with the support of the Interdisciplinary Center for Mathematical and Computational Modelling (ICM) at the University of Warsaw under grant no. GB76-1.}

\appendix

\red{
\section{Equilibration curves} \label{sec:equilib}

\begin{figure*}[htb]
    \centering
    \includegraphics[width=0.4\linewidth]{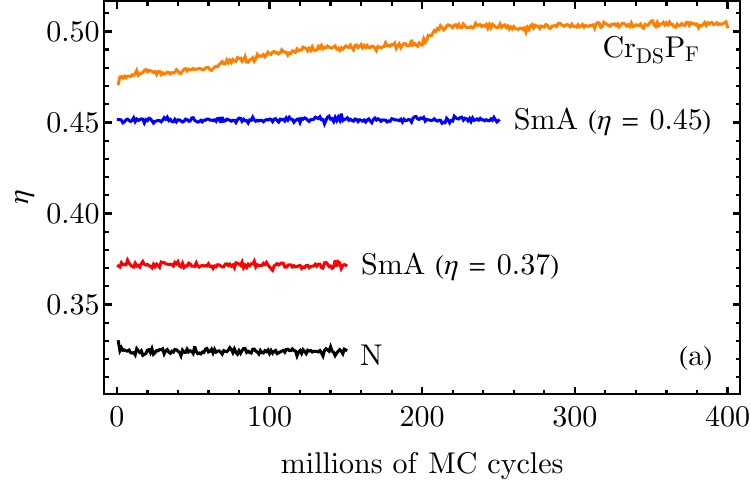}
    \includegraphics[width=0.4\linewidth]{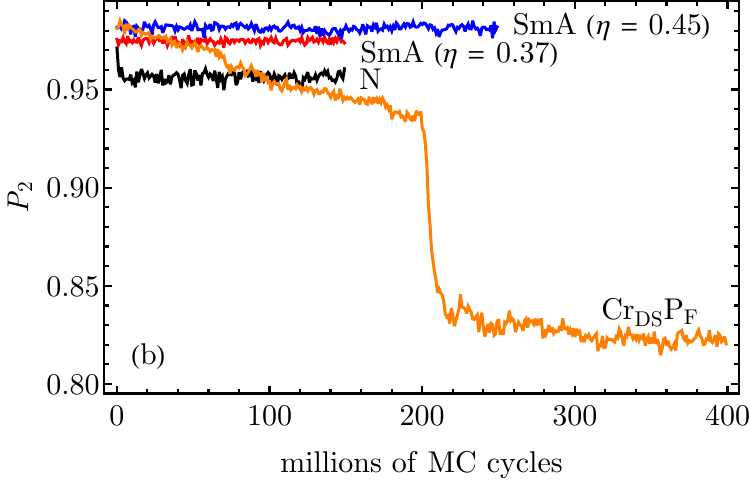}
    \includegraphics[width=0.4\linewidth]{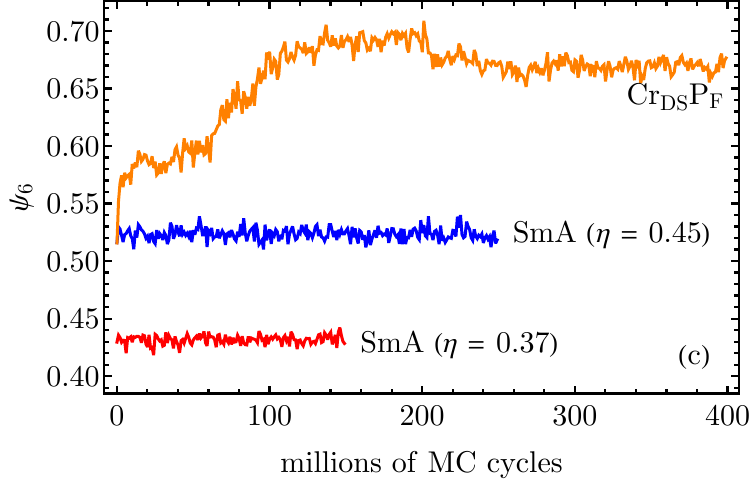}
    \includegraphics[width=0.4\linewidth]{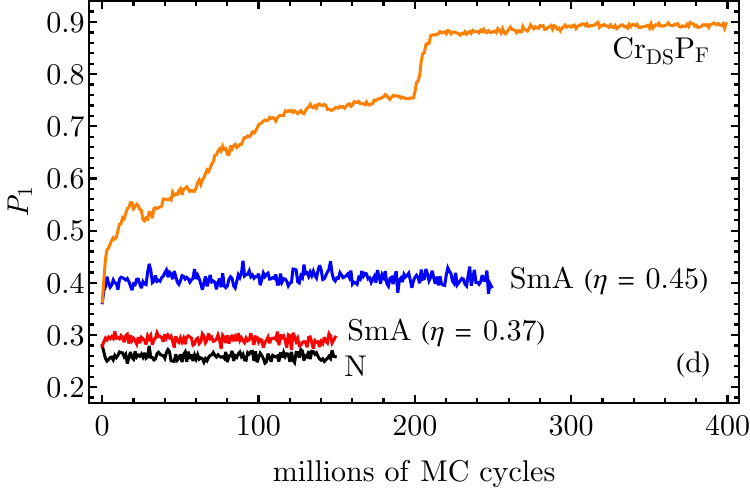}
    \caption{\red{Equilibration curves of selected quantities: (a) packing fraction $\eta$, (b) nematic order parameter $P_2$, (c) hexatic bond order parameter $\psi_6$ and (d) local polarization $P_1$ for N ($\eta = 0.32$), low-density SmA ($\eta = 0.37$), high-density SmA ($\eta = 0.45$) and $\text{Cr}_\text{DS}\text{P}_\text{F}$ ($\eta = 0.51$).}}
    \label{fig:equilib}
\end{figure*}

Fig.~\ref{fig:equilib} shows the equilibration curves for selected quantities: packing fraction $\eta$, nematic order parameter $P_2$ hexatic bond order parameter $\psi_6$ and local polarization $P_1$. For nematics and smectics the system reaches equilibrium relatively fast, not needing more than $2 \cross 10^7$ MC cycles. On the other hand, the crystalline $\text{Cr}_\text{DS}\text{P}_\text{F}$ phase requires significantly more time to develop -- the observables reach their thermal values only after $3 \cross 10^8$ full MC cycles.
}

\bibliography{main.bib} 

\providecommand*{\mcitethebibliography}{\thebibliography}
\csname @ifundefined\endcsname{endmcitethebibliography}
{\let\endmcitethebibliography\endthebibliography}{}
\begin{mcitethebibliography}{56}
\providecommand*{\natexlab}[1]{#1}
\providecommand*{\mciteSetBstSublistMode}[1]{}
\providecommand*{\mciteSetBstMaxWidthForm}[2]{}
\providecommand*{\mciteBstWouldAddEndPuncttrue}
  {\def\EndOfBibitem{\unskip.}}
\providecommand*{\mciteBstWouldAddEndPunctfalse}
  {\let\EndOfBibitem\relax}
\providecommand*{\mciteSetBstMidEndSepPunct}[3]{}
\providecommand*{\mciteSetBstSublistLabelBeginEnd}[3]{}
\providecommand*{\EndOfBibitem}{}
\mciteSetBstSublistMode{f}
\mciteSetBstMaxWidthForm{subitem}
{(\emph{\alph{mcitesubitemcount}})}
\mciteSetBstSublistLabelBeginEnd{\mcitemaxwidthsubitemform\space}
{\relax}{\relax}

\bibitem[Reinitzer(1888)]{Reinitzer1888}
F.~Reinitzer, \emph{Monatshefte f{\"u}r Chemie und verwandte Teile anderer
  Wissenschaften}, 1888, \textbf{9}, 421--441\relax
\mciteBstWouldAddEndPuncttrue
\mciteSetBstMidEndSepPunct{\mcitedefaultmidpunct}
{\mcitedefaultendpunct}{\mcitedefaultseppunct}\relax
\EndOfBibitem
\bibitem[Stephen and Straley(1974)]{Stephen1974}
M.~J. Stephen and J.~P. Straley, \emph{Rev. Mod. Phys.}, 1974, \textbf{46},
  617\relax
\mciteBstWouldAddEndPuncttrue
\mciteSetBstMidEndSepPunct{\mcitedefaultmidpunct}
{\mcitedefaultendpunct}{\mcitedefaultseppunct}\relax
\EndOfBibitem
\bibitem[Chandrasekhar(1992)]{Chandrasekhar1992}
S.~Chandrasekhar, \emph{Liquid Crystals}, Cambridge University Press, 2nd edn,
  1992\relax
\mciteBstWouldAddEndPuncttrue
\mciteSetBstMidEndSepPunct{\mcitedefaultmidpunct}
{\mcitedefaultendpunct}{\mcitedefaultseppunct}\relax
\EndOfBibitem
\bibitem[de~Gennes and Prost(1993)]{deGennes1993}
P.~G. de~Gennes and J.~Prost, \emph{The Physics of Liquid Crystals}, Clarendon
  Press, 2nd edn, 1993\relax
\mciteBstWouldAddEndPuncttrue
\mciteSetBstMidEndSepPunct{\mcitedefaultmidpunct}
{\mcitedefaultendpunct}{\mcitedefaultseppunct}\relax
\EndOfBibitem
\bibitem[Uchida \emph{et~al.}(2022)Uchida, Soberats, Gupta, and
  Kato]{Uchida2022}
J.~Uchida, B.~Soberats, M.~Gupta and T.~Kato, \emph{Adv. Mat.}, 2022,
  \textbf{34}, 2109063\relax
\mciteBstWouldAddEndPuncttrue
\mciteSetBstMidEndSepPunct{\mcitedefaultmidpunct}
{\mcitedefaultendpunct}{\mcitedefaultseppunct}\relax
\EndOfBibitem
\bibitem[Oseen(1933)]{Oseen1933}
C.~W. Oseen, \emph{Trans. Faraday Soc.}, 1933, \textbf{29}, 883--899\relax
\mciteBstWouldAddEndPuncttrue
\mciteSetBstMidEndSepPunct{\mcitedefaultmidpunct}
{\mcitedefaultendpunct}{\mcitedefaultseppunct}\relax
\EndOfBibitem
\bibitem[Zocher(1933)]{Zocher1933}
H.~Zocher, \emph{Trans. Faraday Soc.}, 1933, \textbf{29}, 945--957\relax
\mciteBstWouldAddEndPuncttrue
\mciteSetBstMidEndSepPunct{\mcitedefaultmidpunct}
{\mcitedefaultendpunct}{\mcitedefaultseppunct}\relax
\EndOfBibitem
\bibitem[Frank(1958)]{Frank1958}
F.~C. Frank, \emph{Discuss. Faraday Soc.}, 1958, \textbf{25}, 19--28\relax
\mciteBstWouldAddEndPuncttrue
\mciteSetBstMidEndSepPunct{\mcitedefaultmidpunct}
{\mcitedefaultendpunct}{\mcitedefaultseppunct}\relax
\EndOfBibitem
\bibitem[Demus \emph{et~al.}(2008)Demus, Goodby, Gray, Spiess, and
  Vill]{Demus2008}
D.~Demus, J.~W. Goodby, G.~W. Gray, H.~W. Spiess and V.~Vill, \emph{Handbook of
  Liquid Crystals, Volume 3: High Molecular Weight Liquid Crystals}, John Wiley
  \& Sons, 2008, vol.~3\relax
\mciteBstWouldAddEndPuncttrue
\mciteSetBstMidEndSepPunct{\mcitedefaultmidpunct}
{\mcitedefaultendpunct}{\mcitedefaultseppunct}\relax
\EndOfBibitem
\bibitem[Luckhurst \emph{et~al.}(2001)Luckhurst, Dunmur, and
  Fukuda]{Luckhurst2001}
G.~Luckhurst, D.~A. Dunmur and A.~Fukuda, \emph{Physical properties of liquid
  crystals: nematics}, IET, 2001\relax
\mciteBstWouldAddEndPuncttrue
\mciteSetBstMidEndSepPunct{\mcitedefaultmidpunct}
{\mcitedefaultendpunct}{\mcitedefaultseppunct}\relax
\EndOfBibitem
\bibitem[Meyer(1976)]{Meyer1976}
R.~B. Meyer, \emph{Molecular Fluids}, Gordon and Breach Science Publishers,
  1976, p. 271\relax
\mciteBstWouldAddEndPuncttrue
\mciteSetBstMidEndSepPunct{\mcitedefaultmidpunct}
{\mcitedefaultendpunct}{\mcitedefaultseppunct}\relax
\EndOfBibitem
\bibitem[Dozov(2001)]{Dozov2001}
I.~Dozov, \emph{Europhys. Let. ({EPL})}, 2001, \textbf{56}, 247--253\relax
\mciteBstWouldAddEndPuncttrue
\mciteSetBstMidEndSepPunct{\mcitedefaultmidpunct}
{\mcitedefaultendpunct}{\mcitedefaultseppunct}\relax
\EndOfBibitem
\bibitem[Memmer(2002)]{Memmer2002}
R.~Memmer, \emph{Liq. Cryst.}, 2002, \textbf{29}, 483--496\relax
\mciteBstWouldAddEndPuncttrue
\mciteSetBstMidEndSepPunct{\mcitedefaultmidpunct}
{\mcitedefaultendpunct}{\mcitedefaultseppunct}\relax
\EndOfBibitem
\bibitem[Shamid \emph{et~al.}(2013)Shamid, Dhakal, and Selinger]{Shamid2013}
S.~M. Shamid, S.~Dhakal and J.~V. Selinger, \emph{Phys. Rev. E}, 2013,
  \textbf{87}, 052503\relax
\mciteBstWouldAddEndPuncttrue
\mciteSetBstMidEndSepPunct{\mcitedefaultmidpunct}
{\mcitedefaultendpunct}{\mcitedefaultseppunct}\relax
\EndOfBibitem
\bibitem[Borshch \emph{et~al.}(2013)Borshch, Kim, Xiang, Gao, J{\'a}kli, Panov,
  Vij, Imrie, Tamba, Mehl, and Lavrentovich]{Borshch2013}
V.~Borshch, Y.-K. Kim, J.~Xiang, M.~Gao, A.~J{\'a}kli, V.~P. Panov, J.~K. Vij,
  C.~T. Imrie, M.~G. Tamba, G.~H. Mehl and O.~D. Lavrentovich, \emph{Nat.
  Commun.}, 2013, \textbf{4}, 2635\relax
\mciteBstWouldAddEndPuncttrue
\mciteSetBstMidEndSepPunct{\mcitedefaultmidpunct}
{\mcitedefaultendpunct}{\mcitedefaultseppunct}\relax
\EndOfBibitem
\bibitem[Chen \emph{et~al.}(2013)Chen, Porada, Hooper, Klittnick, Shen,
  Tuchband, Korblova, Bedrov, Walba, Glaser, Maclennan, and Clark]{Chen2013}
D.~Chen, J.~H. Porada, J.~B. Hooper, A.~Klittnick, Y.~Shen, M.~R. Tuchband,
  E.~Korblova, D.~Bedrov, D.~M. Walba, M.~A. Glaser, J.~E. Maclennan and N.~A.
  Clark, \emph{PNAS}, 2013, \textbf{110}, 15931--15936\relax
\mciteBstWouldAddEndPuncttrue
\mciteSetBstMidEndSepPunct{\mcitedefaultmidpunct}
{\mcitedefaultendpunct}{\mcitedefaultseppunct}\relax
\EndOfBibitem
\bibitem[Greco and Ferrarini(2015)]{Greco2015}
C.~Greco and A.~Ferrarini, \emph{Phys. Rev. Let.}, 2015, \textbf{115},
  147801\relax
\mciteBstWouldAddEndPuncttrue
\mciteSetBstMidEndSepPunct{\mcitedefaultmidpunct}
{\mcitedefaultendpunct}{\mcitedefaultseppunct}\relax
\EndOfBibitem
\bibitem[D’Alessandro \emph{et~al.}(2017)D’Alessandro, Luckhurst, and
  Sluckin]{Allesandro2017}
G.~D’Alessandro, G.~R. Luckhurst and T.~J. Sluckin, \emph{Liq. Cryst.}, 2017,
  \textbf{44}, 1--3\relax
\mciteBstWouldAddEndPuncttrue
\mciteSetBstMidEndSepPunct{\mcitedefaultmidpunct}
{\mcitedefaultendpunct}{\mcitedefaultseppunct}\relax
\EndOfBibitem
\bibitem[Chiappini and Dijkstra(2021)]{Chiappini2021}
M.~Chiappini and M.~Dijkstra, \emph{Nat. Commun.}, 2021, \textbf{12},
  2157\relax
\mciteBstWouldAddEndPuncttrue
\mciteSetBstMidEndSepPunct{\mcitedefaultmidpunct}
{\mcitedefaultendpunct}{\mcitedefaultseppunct}\relax
\EndOfBibitem
\bibitem[Kubala \emph{et~al.}(2022)Kubala, Tomczyk, and Cie{\'s}la]{Kubala2022}
P.~Kubala, W.~Tomczyk and M.~Cie{\'s}la, \emph{J. Mol. Liq.}, 2022,
  \textbf{367}, 120156\relax
\mciteBstWouldAddEndPuncttrue
\mciteSetBstMidEndSepPunct{\mcitedefaultmidpunct}
{\mcitedefaultendpunct}{\mcitedefaultseppunct}\relax
\EndOfBibitem
\bibitem[Harris \emph{et~al.}(1999)Harris, Kamien, and Lubensky]{Harris1999}
A.~B. Harris, R.~D. Kamien and T.~C. Lubensky, \emph{Rev. Mod. Phys.}, 1999,
  \textbf{71}, 1745\relax
\mciteBstWouldAddEndPuncttrue
\mciteSetBstMidEndSepPunct{\mcitedefaultmidpunct}
{\mcitedefaultendpunct}{\mcitedefaultseppunct}\relax
\EndOfBibitem
\bibitem[Greco \emph{et~al.}(2014)Greco, Luckhurst, and Ferrarini]{Greco2014}
C.~Greco, G.~R. Luckhurst and A.~Ferrarini, \emph{Soft Mat.}, 2014,
  \textbf{10}, 9318--9323\relax
\mciteBstWouldAddEndPuncttrue
\mciteSetBstMidEndSepPunct{\mcitedefaultmidpunct}
{\mcitedefaultendpunct}{\mcitedefaultseppunct}\relax
\EndOfBibitem
\bibitem[Dhakal and Selinger(2010)]{Dhakal2010}
S.~Dhakal and J.~V. Selinger, \emph{Phys. Rev. E}, 2010, \textbf{81},
  031704\relax
\mciteBstWouldAddEndPuncttrue
\mciteSetBstMidEndSepPunct{\mcitedefaultmidpunct}
{\mcitedefaultendpunct}{\mcitedefaultseppunct}\relax
\EndOfBibitem
\bibitem[Mertelj \emph{et~al.}(2018)Mertelj, Cmok, Sebasti{\'a}n, Mandle,
  Parker, Whitwood, Goodby, and {\v{C}}opi{\v{c}}]{Mertelj2018}
A.~Mertelj, L.~Cmok, N.~Sebasti{\'a}n, R.~J. Mandle, R.~R. Parker, A.~C.
  Whitwood, J.~W. Goodby and M.~{\v{C}}opi{\v{c}}, \emph{Phys. Rev. X}, 2018,
  \textbf{8}, 041025\relax
\mciteBstWouldAddEndPuncttrue
\mciteSetBstMidEndSepPunct{\mcitedefaultmidpunct}
{\mcitedefaultendpunct}{\mcitedefaultseppunct}\relax
\EndOfBibitem
\bibitem[Rosseto and Selinger(2020)]{Rosseto2020}
M.~P. Rosseto and J.~V. Selinger, \emph{Phys. Rev. E}, 2020, \textbf{101},
  052707\relax
\mciteBstWouldAddEndPuncttrue
\mciteSetBstMidEndSepPunct{\mcitedefaultmidpunct}
{\mcitedefaultendpunct}{\mcitedefaultseppunct}\relax
\EndOfBibitem
\bibitem[Sebasti{\'a}n \emph{et~al.}(2020)Sebasti{\'a}n, Cmok, Mandle, de~la
  Fuente, Olenik, {\v{C}}opi{\v{c}}, and Mertelj]{Sebastian2020}
N.~Sebasti{\'a}n, L.~Cmok, R.~J. Mandle, M.~R. de~la Fuente, I.~D. Olenik,
  M.~{\v{C}}opi{\v{c}} and A.~Mertelj, \emph{Phys. Rev. Let.}, 2020,
  \textbf{124}, 037801\relax
\mciteBstWouldAddEndPuncttrue
\mciteSetBstMidEndSepPunct{\mcitedefaultmidpunct}
{\mcitedefaultendpunct}{\mcitedefaultseppunct}\relax
\EndOfBibitem
\bibitem[Archbold \emph{et~al.}(2015)Archbold, Davis, Mandle, Cowling, and
  Goodby]{Archbold2015}
C.~T. Archbold, E.~J. Davis, R.~J. Mandle, S.~J. Cowling and J.~W. Goodby,
  \emph{Soft Mat.}, 2015, \textbf{11}, 7547--7557\relax
\mciteBstWouldAddEndPuncttrue
\mciteSetBstMidEndSepPunct{\mcitedefaultmidpunct}
{\mcitedefaultendpunct}{\mcitedefaultseppunct}\relax
\EndOfBibitem
\bibitem[Paj{\k{a}}k \emph{et~al.}(2018)Paj{\k{a}}k, Longa, and
  Chrzanowska]{Pajak2018}
G.~Paj{\k{a}}k, L.~Longa and A.~Chrzanowska, \emph{PNAS}, 2018, \textbf{115},
  E10303--E10312\relax
\mciteBstWouldAddEndPuncttrue
\mciteSetBstMidEndSepPunct{\mcitedefaultmidpunct}
{\mcitedefaultendpunct}{\mcitedefaultseppunct}\relax
\EndOfBibitem
\bibitem[Chaturvedi and Kamien(2019)]{Chaturvedi2019}
N.~Chaturvedi and R.~D. Kamien, \emph{Phys. Rev. E}, 2019, \textbf{100},
  022704\relax
\mciteBstWouldAddEndPuncttrue
\mciteSetBstMidEndSepPunct{\mcitedefaultmidpunct}
{\mcitedefaultendpunct}{\mcitedefaultseppunct}\relax
\EndOfBibitem
\bibitem[Fern{\'a}ndez-Rico \emph{et~al.}(2020)Fern{\'a}ndez-Rico, Chiappini,
  Yanagishima, de~Sousa, Aarts, Dijkstra, and Dullens]{Fernandez2020}
C.~Fern{\'a}ndez-Rico, M.~Chiappini, T.~Yanagishima, H.~de~Sousa, D.~G. Aarts,
  M.~Dijkstra and R.~P. Dullens, \emph{Science}, 2020, \textbf{369},
  950--955\relax
\mciteBstWouldAddEndPuncttrue
\mciteSetBstMidEndSepPunct{\mcitedefaultmidpunct}
{\mcitedefaultendpunct}{\mcitedefaultseppunct}\relax
\EndOfBibitem
\bibitem[Mandle \emph{et~al.}(2017)Mandle, Cowling, and Goodby]{Mandle2017}
R.~J. Mandle, S.~J. Cowling and J.~W. Goodby, \emph{Chemistry--A European
  Journal}, 2017, \textbf{23}, 14554--14562\relax
\mciteBstWouldAddEndPuncttrue
\mciteSetBstMidEndSepPunct{\mcitedefaultmidpunct}
{\mcitedefaultendpunct}{\mcitedefaultseppunct}\relax
\EndOfBibitem
\bibitem[Chen \emph{et~al.}(2020)Chen, Korblova, Dong, Wei, Shao, Radzihovsky,
  Glaser, Maclennan, Bedrov, Walba,\emph{et~al.}]{Chen2020}
X.~Chen, E.~Korblova, D.~Dong, X.~Wei, R.~Shao, L.~Radzihovsky, M.~A. Glaser,
  J.~E. Maclennan, D.~Bedrov, D.~M. Walba \emph{et~al.}, \emph{PNAS}, 2020,
  \textbf{117}, 14021--14031\relax
\mciteBstWouldAddEndPuncttrue
\mciteSetBstMidEndSepPunct{\mcitedefaultmidpunct}
{\mcitedefaultendpunct}{\mcitedefaultseppunct}\relax
\EndOfBibitem
\bibitem[Sebasti{\'a}n \emph{et~al.}(2022)Sebasti{\'a}n, {\v{C}}opi{\v{c}}, and
  Mertelj]{Sebastian2022}
N.~Sebasti{\'a}n, M.~{\v{C}}opi{\v{c}} and A.~Mertelj, \emph{Phys. Rev. E},
  2022, \textbf{106}, 021001\relax
\mciteBstWouldAddEndPuncttrue
\mciteSetBstMidEndSepPunct{\mcitedefaultmidpunct}
{\mcitedefaultendpunct}{\mcitedefaultseppunct}\relax
\EndOfBibitem
\bibitem[De~Gregorio \emph{et~al.}(2016)De~Gregorio, Frezza, Greco, and
  Ferrarini]{Gregorio2016}
P.~De~Gregorio, E.~Frezza, C.~Greco and A.~Ferrarini, \emph{Soft Mat.}, 2016,
  \textbf{12}, 5188--5198\relax
\mciteBstWouldAddEndPuncttrue
\mciteSetBstMidEndSepPunct{\mcitedefaultmidpunct}
{\mcitedefaultendpunct}{\mcitedefaultseppunct}\relax
\EndOfBibitem
\bibitem[Gay and Berne(1981)]{Gay1981}
J.~Gay and B.~Berne, \emph{J. Chem. Phys.}, 1981, \textbf{74}, 3316--3319\relax
\mciteBstWouldAddEndPuncttrue
\mciteSetBstMidEndSepPunct{\mcitedefaultmidpunct}
{\mcitedefaultendpunct}{\mcitedefaultseppunct}\relax
\EndOfBibitem
\bibitem[Berardi \emph{et~al.}(2001)Berardi, Ricci, and Zannoni]{Berardi2001}
R.~Berardi, M.~Ricci and C.~Zannoni, \emph{Chem. Phys. Chem.}, 2001,
  \textbf{2}, 443--447\relax
\mciteBstWouldAddEndPuncttrue
\mciteSetBstMidEndSepPunct{\mcitedefaultmidpunct}
{\mcitedefaultendpunct}{\mcitedefaultseppunct}\relax
\EndOfBibitem
\bibitem[Barmes \emph{et~al.}(2003)Barmes, Ricci, Zannoni, and
  Cleaver]{Barmes2003}
F.~Barmes, M.~Ricci, C.~Zannoni and D.~Cleaver, \emph{Phys. Rev. E}, 2003,
  \textbf{68}, 021708\relax
\mciteBstWouldAddEndPuncttrue
\mciteSetBstMidEndSepPunct{\mcitedefaultmidpunct}
{\mcitedefaultendpunct}{\mcitedefaultseppunct}\relax
\EndOfBibitem
\bibitem[Houssa \emph{et~al.}(2009)Houssa, Rull, and
  Romero-Enrique]{Houssa2009}
M.~Houssa, L.~F. Rull and J.~M. Romero-Enrique, \emph{J. Chem. Phys.}, 2009,
  \textbf{130}, 154504\relax
\mciteBstWouldAddEndPuncttrue
\mciteSetBstMidEndSepPunct{\mcitedefaultmidpunct}
{\mcitedefaultendpunct}{\mcitedefaultseppunct}\relax
\EndOfBibitem
\bibitem[Kubala \emph{et~al.}(2022)Kubala, Cie{\'s}la, and
  Longa]{Kubala2022phases}
P.~Kubala, M.~Cie{\'s}la and L.~Longa, \emph{arXiv preprint arXiv:2210.04737},
  2022\relax
\mciteBstWouldAddEndPuncttrue
\mciteSetBstMidEndSepPunct{\mcitedefaultmidpunct}
{\mcitedefaultendpunct}{\mcitedefaultseppunct}\relax
\EndOfBibitem
\bibitem[Wood(1968)]{Wood1968}
W.~W. Wood, \emph{Physics of simple liquids}, North-Holland, 1968\relax
\mciteBstWouldAddEndPuncttrue
\mciteSetBstMidEndSepPunct{\mcitedefaultmidpunct}
{\mcitedefaultendpunct}{\mcitedefaultseppunct}\relax
\EndOfBibitem
\bibitem[Allen and Tildesley(2017)]{Allen2017}
M.~P. Allen and D.~J. Tildesley, \emph{Computer simulation of liquids}, Oxford
  university press, 2017\relax
\mciteBstWouldAddEndPuncttrue
\mciteSetBstMidEndSepPunct{\mcitedefaultmidpunct}
{\mcitedefaultendpunct}{\mcitedefaultseppunct}\relax
\EndOfBibitem
\bibitem[Allen(2019)]{Allen2019}
M.~P. Allen, \emph{Mol. Phys.}, 2019, \textbf{117}, 2391--2417\relax
\mciteBstWouldAddEndPuncttrue
\mciteSetBstMidEndSepPunct{\mcitedefaultmidpunct}
{\mcitedefaultendpunct}{\mcitedefaultseppunct}\relax
\EndOfBibitem
\bibitem[Wood(1968)]{Wood1968jcp}
W.~Wood, \emph{J. Chem. Phys.}, 1968, \textbf{48}, 415--434\relax
\mciteBstWouldAddEndPuncttrue
\mciteSetBstMidEndSepPunct{\mcitedefaultmidpunct}
{\mcitedefaultendpunct}{\mcitedefaultseppunct}\relax
\EndOfBibitem
\bibitem[Vieillard-Baron(1974)]{vieillard1974}
J.~Vieillard-Baron, \emph{Mol. Phys.}, 1974, \textbf{28}, 809--818\relax
\mciteBstWouldAddEndPuncttrue
\mciteSetBstMidEndSepPunct{\mcitedefaultmidpunct}
{\mcitedefaultendpunct}{\mcitedefaultseppunct}\relax
\EndOfBibitem
\bibitem[Eppenga and Frenkel(1984)]{Eppenga1984}
R.~Eppenga and D.~Frenkel, \emph{Mol. Phys.}, 1984, \textbf{52},
  1303--1334\relax
\mciteBstWouldAddEndPuncttrue
\mciteSetBstMidEndSepPunct{\mcitedefaultmidpunct}
{\mcitedefaultendpunct}{\mcitedefaultseppunct}\relax
\EndOfBibitem
\bibitem[Kittel and McEuen(2018)]{Kittel2018}
C.~Kittel and P.~McEuen, \emph{Introduction to solid state physics}, John Wiley
  \& Sons, 2018\relax
\mciteBstWouldAddEndPuncttrue
\mciteSetBstMidEndSepPunct{\mcitedefaultmidpunct}
{\mcitedefaultendpunct}{\mcitedefaultseppunct}\relax
\EndOfBibitem
\bibitem[Nelson(2012)]{Nelson2012}
D.~Nelson, \emph{Bond-orientational order in condensed matter systems},
  Springer Science \& Business Media, 2012\relax
\mciteBstWouldAddEndPuncttrue
\mciteSetBstMidEndSepPunct{\mcitedefaultmidpunct}
{\mcitedefaultendpunct}{\mcitedefaultseppunct}\relax
\EndOfBibitem
\bibitem[Stone(1978)]{Stone1978}
A.~J. Stone, \emph{Mol. Phys.}, 1978, \textbf{36}, 241--256\relax
\mciteBstWouldAddEndPuncttrue
\mciteSetBstMidEndSepPunct{\mcitedefaultmidpunct}
{\mcitedefaultendpunct}{\mcitedefaultseppunct}\relax
\EndOfBibitem
\bibitem[Chandrasekhar and Madhusudana(1980)]{Chandrasekhar1980}
S.~Chandrasekhar and N.~V. Madhusudana, \emph{Annu. Rev. Mater. Sci.}, 1980,
  \textbf{10}, 133--155\relax
\mciteBstWouldAddEndPuncttrue
\mciteSetBstMidEndSepPunct{\mcitedefaultmidpunct}
{\mcitedefaultendpunct}{\mcitedefaultseppunct}\relax
\EndOfBibitem
\bibitem[Vega \emph{et~al.}(2001)Vega, McBride, and Macdowell]{Vega2001}
C.~Vega, C.~McBride and L.~G. Macdowell, \emph{J. Chem. Phys.}, 2001,
  \textbf{115}, 4203--4211\relax
\mciteBstWouldAddEndPuncttrue
\mciteSetBstMidEndSepPunct{\mcitedefaultmidpunct}
{\mcitedefaultendpunct}{\mcitedefaultseppunct}\relax
\EndOfBibitem
\bibitem[Lansac \emph{et~al.}(2003)Lansac, Maiti, Clark, and
  Glaser]{Lansac2003}
Y.~Lansac, P.~K. Maiti, N.~A. Clark and M.~A. Glaser, \emph{Phys. Rev. E},
  2003, \textbf{67}, 011703\relax
\mciteBstWouldAddEndPuncttrue
\mciteSetBstMidEndSepPunct{\mcitedefaultmidpunct}
{\mcitedefaultendpunct}{\mcitedefaultseppunct}\relax
\EndOfBibitem
\bibitem[Singh(2000)]{Singh2000}
S.~Singh, \emph{Phys. Rep.}, 2000, \textbf{324}, 107--269\relax
\mciteBstWouldAddEndPuncttrue
\mciteSetBstMidEndSepPunct{\mcitedefaultmidpunct}
{\mcitedefaultendpunct}{\mcitedefaultseppunct}\relax
\EndOfBibitem
\bibitem[McGrother \emph{et~al.}(1996)McGrother, Williamson, and
  Jackson]{McGrother1996}
S.~C. McGrother, D.~C. Williamson and G.~Jackson, \emph{J. Chem. Phys.}, 1996,
  \textbf{104}, 6755--6771\relax
\mciteBstWouldAddEndPuncttrue
\mciteSetBstMidEndSepPunct{\mcitedefaultmidpunct}
{\mcitedefaultendpunct}{\mcitedefaultseppunct}\relax
\EndOfBibitem
\bibitem[Polson and Frenkel(1997)]{Polson1997}
J.~M. Polson and D.~Frenkel, \emph{Phys. Rev. E}, 1997, \textbf{56},
  R6260\relax
\mciteBstWouldAddEndPuncttrue
\mciteSetBstMidEndSepPunct{\mcitedefaultmidpunct}
{\mcitedefaultendpunct}{\mcitedefaultseppunct}\relax
\EndOfBibitem
\bibitem[Parsons(1979)]{Parsons1979}
J.~Parsons, \emph{Phys. Rev. A}, 1979, \textbf{19}, 1225\relax
\mciteBstWouldAddEndPuncttrue
\mciteSetBstMidEndSepPunct{\mcitedefaultmidpunct}
{\mcitedefaultendpunct}{\mcitedefaultseppunct}\relax
\EndOfBibitem
\bibitem[Lee(1987)]{Lee1987}
S.-D. Lee, \emph{J. Chem. Phys}, 1987, \textbf{87}, 4972--4974\relax
\mciteBstWouldAddEndPuncttrue
\mciteSetBstMidEndSepPunct{\mcitedefaultmidpunct}
{\mcitedefaultendpunct}{\mcitedefaultseppunct}\relax
\EndOfBibitem
\end{mcitethebibliography}
\bibliographystyle{rsc} 

\end{document}